\documentclass[utf8]{frontiersSCNS} 

\usepackage{url,hyperref,microtype,subcaption}
\usepackage[onehalfspacing]{setspace}

\usepackage[]{natbib}
\usepackage{amsfonts}
\usepackage{amsmath}
\usepackage{amssymb}
\usepackage{mathtools}
\usepackage{mathabx} 
\usepackage{xspace} 
\usepackage[normalem]{ulem}

\newcommand\bb[1] {   \mbox{\boldmath{$#1$}}  }

\newcommand{\BV}{Brunt-V\"ais\"al\"a\ }
\newcommand\er{ \bb{e_r}  }

\newcommand\ylm{ Y_\ell^m}

\newcommand{\Msun}{{\rm M}_\odot\xspace}

\newcommand{\ltilde}{\widetilde{\ell}\xspace}
\newcommand{\ntilde}{\widetilde{n}\xspace}

\newcommand{\kms}{km s$^{-1}$}

\newcommand\xoutpars[1]{\let\helpcmd\xout\parhelp#1\par\relax\relax}
\newcommand\soutpars[1]{\let\helpcmd\sout\parhelp#1\par\relax\relax}
\long\def\parhelp#1\par#2\relax{%
  \helpcmd{#1}\ifx\relax#2\else\par\parhelp#2\relax\fi%
  }

\graphicspath{{jpg_figures/}}

\def\keyFont{\fontsize{8}{11}\helveticabold }
\def\firstAuthorLast{Mirouh} 
\def\Authors{Giovanni M. Mirouh\,$^{1,*}$} 
\def\Address{$^{1}$ Física Teórica y del Cosmos Department, Universidad de Granada, Spain}
\def\corrAuthor{Corresponding Author}
\def\corrEmail{gmm@ugr.es}

\begin{document}
\onecolumn
\firstpage{1}

\title[Mode identification in rapidly-rotating stars]{
Forward modelling and the quest for mode identification in rapidly-rotating stars } 

\author[\firstAuthorLast ]{\Authors} 
\address{} 
\correspondance{} 

\extraAuth{}

\maketitle

\begin{abstract}
  Asteroseismology has opened a window on the internal physics of thousands of
  stars, by relating oscillation spectra properties to the internal physics of
  stars. Mode identification, namely the process of associating a measured
  oscillation frequency to the corresponding mode geometry and properties, is
  the cornerstone of this analysis of seismic spectra. In  rapidly rotating
  stars this identification is a challenging task that remains incomplete, as
  modes assume complex geometries and regular patterns in frequencies get
  scrambled under the influence of the Coriolis force and centrifugal
  flattening.

In this article, I will first discuss the various classes of mode geometries
  that emerge in rapidly-rotating stars and the related frequency and period
  patterns, as predicted by ray dynamics, complete (non-)adiabatic
  calculations, or using the traditional approximation of rotation.  These
  patterns scale with structural quantities and help us derive crucial
  constraints on the structure and evolution of these stars.  I will summarize
  the amazing progress accomplished over the last few years for the deciphering
  of gravity-mode pulsator oscillation spectra, and recent developments based
  on machine-learning classification techniques to distinguish oscillation
  modes and pattern analysis strategies that let us access the underlying
  physics of pressure-mode pulsators. These approaches pave the way to ensemble
  asteroseismology of classical pulsators.

Finally, I will highlight how these recent progress can be combined to improve
  forward seismic modelling. I will focus on the example of Rasalhague, a
  well-known rapid rotator, to illustrate the process and the needed advances
  to obtain \`a-la-carte modelling of such stars.

%

\tiny
 \keyFont{ \section{Keywords:} stars: oscillations, 
                               stars: rotation, 
                               stars: interiors, 
                               stars: evolution, 
                               stars: individual: $\alpha$ Ophiuchi} 
\end{abstract}

\section{Introduction}
\label{sec:intro}

Over the last few decades, helio- and asteroseismology have started a golden
age for stellar physics. The advent of space-based photometry missions (such as
CoRoT, \citealt{corot}, Kepler, \citealt{kepler}, TESS, \citealt{tess}, and
BRITE, \citealt{brite}) has led to the detection of variability in numerous
stars, the measurement of oscillation frequencies and their regular spacings,
and the identification of the mode geometries. This has triggered a cascade of
new theoretical developments for stellar models, oscillation codes and
inversion techniques. 

\begin{figure}
\centerline{ 
  \includegraphics[width=0.33\textwidth]{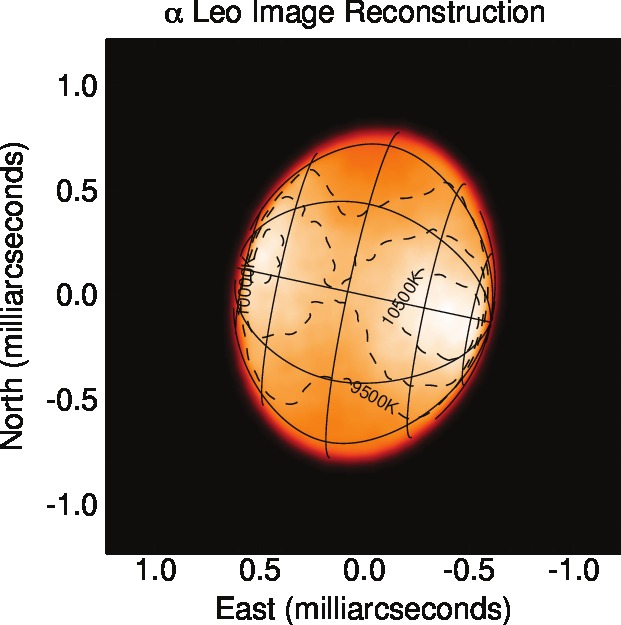}
  \includegraphics[width=0.33\textwidth]{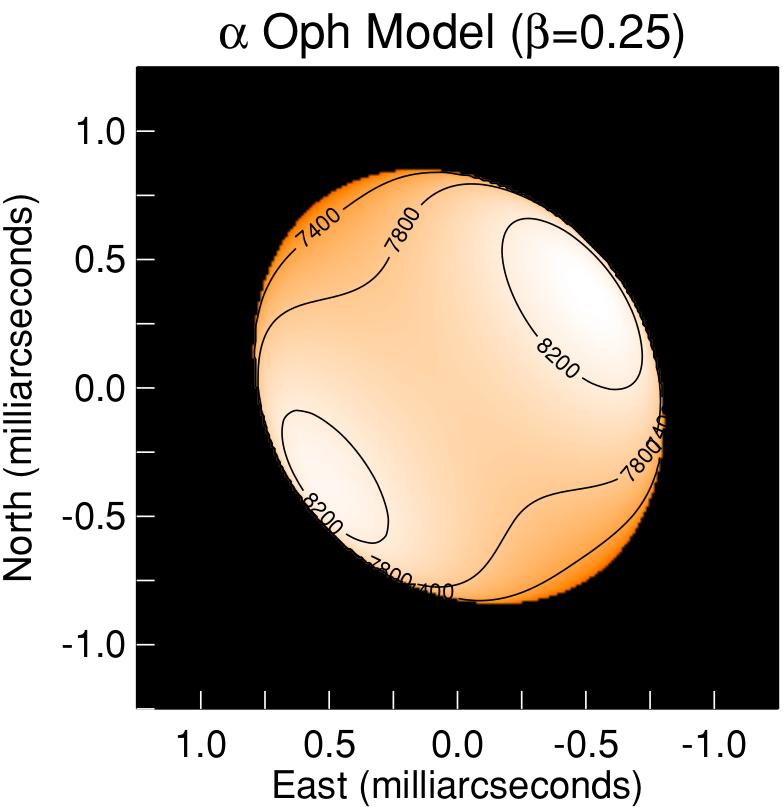}
  \includegraphics[width=0.33\textwidth]{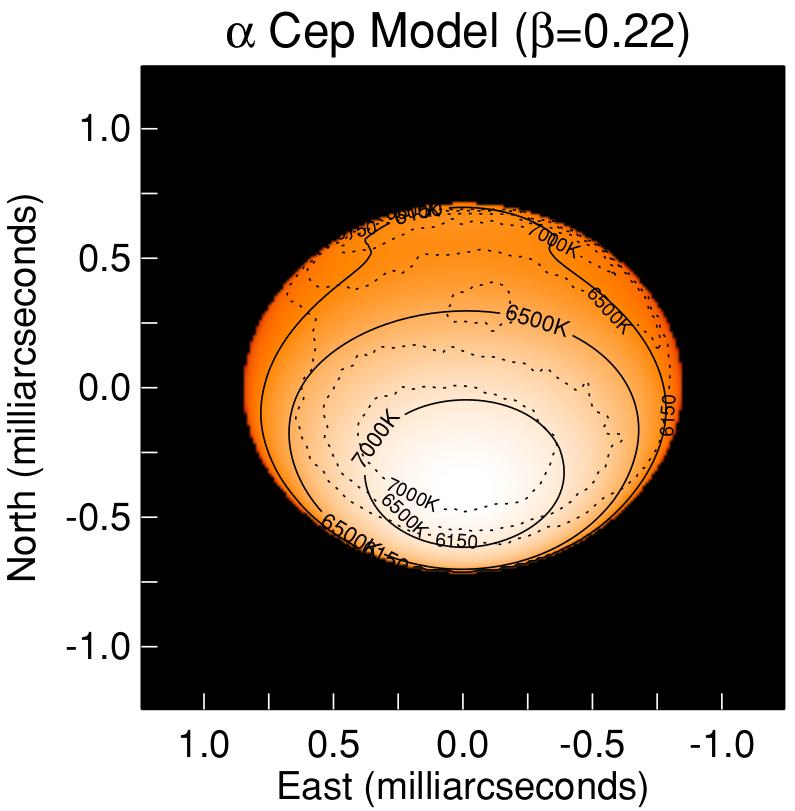}
  }
  \caption{Interferometric observations of rapidly-rotating stars, showing the centrifugal distortion and gravity darkening phenomena. 
  Left to right: Regulus \citep[$\alpha\,$Leo,][]{che_etal11}, Rasalhague \citep[$\alpha\,$Oph,][]{zhao_etal09}, Alderamin \citep[$\alpha\,$Cep,][]{zhao_etal09}.
\label{fig:interfero}}
\end{figure}

However, these developments were long limited to slowly-rotating solar-like and
red giant stars.  Upper main-sequence pulsators present a rapid rotation
\citep{royer2009} that affects both their structure and their oscillations. A
rotating star is centrifugally distorted \citep{potter12}, which induces a
meridional circulation \citep{zahn92, R06} and latitudinal surface temperature
gradients \citep{vonzeipel24, ELR11}.  Figure~\ref{fig:interfero} presents
three examples of interferometric measurements of the surface shape and
temperatures for three close pulsating A stars.

Pulsations can be either pressure or gravito-inertial modes.  Pressure modes
(or acoustic modes) are restored by pressure forces and tend to propagate at
high frequencies. Gravity modes are restored by buoyancy and exist at low
frequencies. Mixed modes can appear when a pressure and a gravity mode
propagating in two separate cavities inside the star at the same frequency
couple. In the presence of rotation, the Coriolis force can act as a restoring
force to create the so-called inertial modes, or combine with buoyancy to
create gravito-inertial modes. At high rotation rates, the Coriolis force is
the dominating effect of rotation on gravity modes while pressure modes are
mostly affected by the centrifugal distortion, both effects modify the
oscillation frequencies and obfuscate regular patterns in the spectra.  This
creates new complex mode geometries \citep{BLR13,LG09,RLR08} that render mode
identification challenging.  Taking these effects fully into account requires
dedicated two-dimensional models \citep{ELR13,RELP2016} and oscillation
calculations \citep{reese_etal09,reese_etal21,ouazzani_etal12}. The
mathematical details of these tools are provided in Reese (this volume).

In this review, I consider pulsating stars on the upper main sequence, that can
be split into four classes: in increasing order of mass, these are the $\gamma\
$Doradus (F-late A), $\delta\ $Scuti (A-F), slowly-pulsating B (SPB, B3-B9) and
$\beta\ $Cephei (B0-B2) stars \citep{aerts_etal10}.  
$\gamma$Dor stars harbour low-frequency gravito-inertial and inertial modes
excited through a combination of convective flux blocking at the base of the
surface convection zone \citep{pesnell87, GKBCN00, DGGGS04, dupret05} in colder
$\gamma$Dor stars and $\kappa$ mechanism in their warmer counterparts
\citep{xiong2016}.
$\delta$Sct stars pulsate at higher frequencies because of acoustic modes
excited by the $\kappa$ mechanism \citep{zhevakin63,DGGGS04}.  These two
classes of modes have more massive counterparts: 
SPBs pulsate with low-frequency (gravito-)inertial modes and
$\beta$Cep stars harbour acoustic and low-order gravity modes, that are
explained in both cases by the $\kappa$ mechanism activated by the metal
opacity bump \citep{cox92,DMP93}.

Due to their overlapping instability ranges, many $\gamma$Dor/$\delta$Sct
hybrids \citep[e.g.][]{handler02, grig10, BDP2015} and SPB/$\beta$Cep hybrids
\citep[e.g.][]{handler04,decat07, handler09, burssens20} have been detected.  A
series of works by \citet{osaki74, LS20, lee22} also proposes resonant coupling
with overstable convection in the core as a possible excitation mechanism for
gravity modes in the envelope of early-type stars, with frequencies close to
the core rotation rate. This may account for rotational-modulation-like signals
observed in SPB stars and hybrid pulsators.

The individual identification of modes in rapidly-rotating stars is hampered by
a limited knowledge of the mode selection mechanisms, the impact of rotation on
both their structure and oscillations, the sheer number of possible oscillation
modes theoretically predicted by high-resolution calculations and our
impossibility to predict mode amplitudes.  Patterns in frequency or period have
opened a window on the internal physics of solar-like and red-giant stars,
after being described theoretically \citep[e.g.][]{shib79,tass80,miglio08} and
identified in observations \citep[such as][]{mosser2013, vrard2016}.  Such
patterns, albeit far more elusive, exist in rapidly-rotating stars and have
been investigated by theoretical works for both p-mode \citep[e.g.][]{LG09,
suarez14, mirouh_etal19} and g-mode spectra \citep[e.g.][]{bouabid2013,
ouazz2017, ouazzani20, dhouib21, TT22}.  Detections in both p-mode
\citep[e.g.][]{AGH15,paparo16a, paparo16b,AGH17,bedding20} and g-mode pulsators
\citep[e.g][]{vanreeth2015,li2020a, pedersen21, garcia22} yield estimates of
the fundamental parameters and the internal physics of an ever-increasing
number of stars.\\
Relating frequency and period spacings to structure quantities thus offers
crucial statistical information and precious constraints for forward modelling
and paves the way to ensemble seismology for classical pulsators
\citep{michel17, bowman18}.

Tools and strategies that are well established for slow rotators have been
redeveloped to take into account rotational effects. One such tool is
line-profile variations, which can yield oscillation frequencies and a partial
identification \citep[see][]{LS90,TS97b, zima06} for both pressure and
gravito-inertial nonradial modes from time series of spectroscopic measurements
\citep[e.g.][]{aerts92, zima06b, shutt21}.  \citet{reese_etal17b} offers a
first adaptation of the technique to complete calculations of pressure modes in
rapid rotators.

In this review I will first describe the impact of the Coriolis force and the
centrifugal flattening of rotating stars on acoustic, inertial and
gravito-inertial mode geometries (section~\ref{sec:geometries}).  Comparing
with non-rotating stars, I will discuss the intricate frequency and period
patterns that rapid-rotator oscillation spectra describe, and the strategies to
extract them from space-based photometry data (section~\ref{sec:patterns}).  I
will then explain how spectroscopic line-profile variations can be evaluated to
derive frequencies of rapid rotators (section~\ref{sec:lpv}), and describe the
example of the forward modelling of Rasalhague (section~\ref{sec:rasalhague}).\\
As I focus on rotation and complete two-dimensional calculations of models and
oscillation modes, I will leave out magnetic fields and their impact on
oscillations. I refer the reader to the literature regarding detections of
magnetic fields in A stars, theoretical study of magneto-acoustic oscillations
and extensive results for rapidly-oscillating Ap (roAp) stars \citep[][and
references therein]{mathis21,hold21}. Interactions between magnetic fields and
low-frequency oscillations have been the focus of a lot of recent work,
regarding internal gravity waves \citep[e.g.]{rogers10,rogers11,leco17,leco22},
coherent gravito-inertial modes \citep[e.g.][]{loi20,prat20,dhouib21} and mixed
modes \citep{loi18, bugnet21} such as depressed dipole modes in red giants
\citep[e.g.][]{garcia14, stello16}.

\section{Mode geometries in rapidly-rotating stars}
\label{sec:geometries}

Contrary to the non-rotating case, oscillations in rapidly-rotating stars
cannot be described using simple spherical harmonics.  Perturbative approaches
quickly show their limits, and strategies that include rotation from the start
(even under some approximations) are necessary to compute models and
oscillations. Several approaches have been implemented, with various levels of
accuracy and complexity. 

\subsection{Including rotation to compute oscillations}

The simplest way of including the effects of rotation is to treat its effects
as corrections to non-rotating solutions of pulsations.  These
corrections are described, for both the centrifugal flattening and the Coriolis
acceleration, as successive powers of the rotation rate $\Omega$, which is
supposed to be small \citep[e.g.][]{ledoux51,GT90,LeeB95,SGD98}.  For high
enough rotation rates, even the highest-order perturbative calculations
misestimate frequencies and describe the mode geometries poorly: \citet{BLR13}
and \citet{reese_thesis} have evaluated the range of applicability of
perturbative treatments for gravito-inertial and pressure modes, respectively.

The traditional approximation \citep{eckart60, berthomieu78} consists in
neglecting the latitudinal component of the rotation vector $\bb{\Omega}$ in
the Coriolis acceleration, assuming $\bb{\Omega} = \Omega \cos\theta\er$.  If
the centrifugal distortion of the star and the perturbation to the
gravitational potential are neglected as well, the system ruling adiabatic
oscillation becomes separable in radius $r$ and colatitude $\theta$,
effectively preventing coupling between modes described by different spherical
harmonics \citep{bildsten96}.  This makes the equations much more
tractable with the added benefit that the horizontal component of the
oscillations can be described mathematically by means of the known Hough
functions \citep{LS87}.  Recent developments by \citet{henneco21,dhouib21}
generalise the traditional approximation to include the effects of the
centrifugal distortion and a differential rotation and predict detectable
impacts on both low- and high-order gravito-inertial modes.  However, and
contrary to the other techniques presented here, the traditional approximation
is a non-perturbative approach only applicable to high-order g
modes (such as those detected in $\gamma$Dor and SPB stars). 

Ray theory relies on the short-wavelength approximation \citep{Gough93, LG09}.
In this approximation, it is possible to combine the oscillation equations into
only one eikonal equation that relates the wavenumber and the location of a ray
in the star \citep{ott93}.  This is akin to deriving geometric optics from wave
optics.  We can thus compute rays, that are the paths waves are expected to
follow inside the star.  While this calculation does not immediately provide
the modes' energy distribution, it provides insights on mode propagation
domains, geometries and identification.

Finally, the most accurate approach consists in the computation of fully
two-dimensional models and the linear calculation of their adiabatic and/or
non-adiabatic oscillations.  This approach provides exact mode geometries and
frequencies, allowing us to search for frequency and period patterns, compute
visibilities and attempt mode-by-mode identification. However, linear mode
calculations cannot yield mode amplitudes, so assumptions (such as the
equirepartition of energy, or the same amplitude for all modes) are necessary
to discuss mode detectability meaningfully.  The tools for these calculations
are complex, and were the focus of numerous efforts summarized in Reese (this
volume). The analysis of the results they yield is challenging as there is no
mathematical function to describe the geometries obtained with this method and
that are presented in the rest of this section.

\subsection{Complete calculations}
The oscillation equations are derived from perturbing the continuity, Euler's
and Poisson's equations (with additional closure equations such as the
equations of state). These equations are presented in detail in the review by
Reese (this volume).  Choosing the geometrical description and discretization
carefully, the perturbed equations form an eigenvalue problem in which each
eigenvalue is an oscillation angular frequency $\omega$, where $\omega =
2\pi\nu$, with the associated eigenvector describing the mode geometry.  In the
non-rotating case, the radial component of the modes is characterised by its
radial order $n$ -- $n$ being a relative integer, the sum of the numbers of
p-mode nodes counted as positive and g-mode nodes counted as negative -- and
the horizontal component is described using a single spherical harmonic $\ylm$
-- where $\ell$ is its degree and $m$ is the azimuthal order.  Without
rotation, all modes at a given $n$ and $\ell$ but different $m$ have the same
frequency.

The equations change when including rotation. In all three of them, the
eigenvalue is shifted in the co-rotating frame, so that the inertial angular
frequency $\omega$ is replaced by the co-rotating value $\omega+m\Omega$. The
definition of a co-rotating frame here is improper but used for simplicity, as
it implies that the rotation is solid: the discussion in this article includes
differential rotation.  Non-radial modes thus split into multiplets as $m$
values can go from $-\ell$ to $\ell$, $m<0$ and $m>0$ being prograde and
retrograde modes, respectively.\\ The Euler equation becomes significantly more
complex, as to include the Coriolis acceleration, centripetal entrainment and
the baroclinicity terms.  These rotational terms create a coupling between
successive spherical harmonic coordinates, so that each mode can no longer be
described as a single spherical harmonic but a (finite) series of them.  No
simple closure relation limits the number of spherical harmonics to use, and
convergence of the solutions requires special care.\\ 
Finally, the boundary conditions also change to accomodate the centrifugal
distortion of the stellar surface.  As the star remains axisymmetric, the
solutions do not depend on the azimuth and $m$ remains separable.  The detailed
derivation of the equations can be found, for instance, in \citet{RLR08,
reese_etal21}.

These equations will yield various mode geometries. I present here the various
subclasses of modes and their properties, as derived from using the traditional
approximation (for g~modes), ray dynamics, and/or complete calculations.

\subsection{Acoustic modes} 
\begin{figure}
\centerline{ \includegraphics[width=\textwidth]{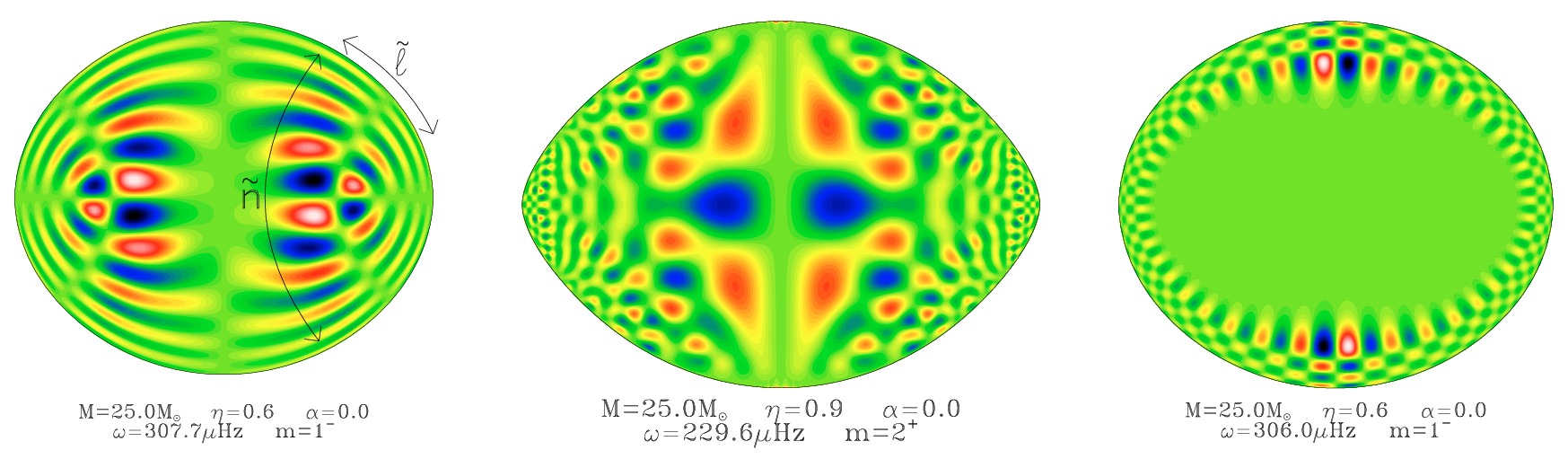}}
  \caption{Different classes of acoustic modes. Left to right: a 2-period
  island mode, a chaotic mode, a whispering gallery mode.  The shown quantity
  is the pressure perturbation rescaled by the square root of the stellar
  density, $\delta P/\sqrt{\rho}$. Figure taken from \citet{reese_etal09}.
\label{fig:geom_p}}
\end{figure}

Pressure (acoustic) modes propagate at high frequencies and usually probe the
outer layers of the stars. They are restored by pressure forces, their
propagation properties are impacted not only by the Coriolis force but also by
the centrifugal distortion.  They are divided into four geometries derived from
ray theory, first put forward by \citet{LG09}: 2-period island, 6-period
island, whispering gallery and chaotic modes.  Figure~\ref{fig:geom_p} shows
examples for the three most-studied classes of acoustic modes.

2-period island modes are the rapidly-rotating counterpart to modes with a low
$(\ell-|m|)$ value -- that is, few latitudinal nodal lines.  At rotation rates
higher than $\sim$40\% of the critical rotation, they propagate in a torus
located around the stellar equator, which corresponds to a short, stable
trajectory in the ray dynamics.  This torus and the associated ray trajectory
are smaller and closer to the equator for higher rotation rates.  It is
possible to define new quantum numbers along the ray trajectory, by defining
$\ntilde$ and $\ltilde$ as the number of nodes along and perpendicular to the
ray trajectory \citep[][and left panel of 
figure~\ref{fig:geom_p}]{reese_etal09}. These quantum numbers can be related to
their non-rotating counterparts through
\begin{equation}
    \ntilde = 2n +\epsilon \quad {\rm and} 
    \quad \ltilde = \frac{\ell-|m|-\epsilon}{2},
    \quad {\rm with} \quad \epsilon = \ell + m \mod 2.
\end{equation}
This mathematical description of the modes is essential to investigate regular
patterns in oscillation spectra, as highlighted in section~\ref{sec:patterns}.
Because they have large lengthscales, these modes induce significant surface
signatures, are expected to be the most visible \citep{reese_etal13,
reese_etal17}, and have been studied extensively
\citep[e.g.][]{reese_etal09, pasek11, ouazz2015, mirouh_etal19,
reese_etal21}.

6-period island modes are relatively rare and have low visibility. Their
geometry is somewhat similar to that of the more common and visible chaotic
modes, described below. Because of this, they are often left out of analyses
(so much so that 2-period island modes are often simply dubbed ``island
modes''). They are the rapidly-rotating counterpart of modes with medium
$(\ell-|m|)$ values.

Chaotic modes emerge from the constructive interference of rays that present a
chaotic trajectory. They also are related to modes with medium $(\ell-|m|)$
values.  \citet{evano_etal19} offers the only extensive study of this
class of modes.  They characterize their geometry by its complexity and
apparent ramdomness. Chaotic modes present irregular nodal lines, except near
the stellar surface where nodal lines tend to become radial.  They are
particularly interesting as they are the only p-mode class to propagate in the
entire stellar interior, probing deep layers while reaching the surface.

Finally, whispering gallery modes propagate close to the surface without
probing deep layers at any point. While their geometry is somewhat affected by
the rotational distortion of the star, they keep lengthscales and a global
structure similar to those of their non-rotating counterparts,
high-$(\ell-|m|)$ modes. As such, the geometry of these modes can be described
with the number of nodal lines crossed at a given depth $d$ (that is, at a
radius $r = R_s (\theta)-d$ where $R_s(\theta)$ is the surface radius, where
$d$ is chosen so that $r$ passes in between the mode's nodal lines) and
perpendicularly to that radius (equivalent to $\ntilde$ and $\ltilde$,
respectively).

\subsection{Gravito-inertial modes}
\begin{figure}
\centerline{ \includegraphics[width=\textwidth]{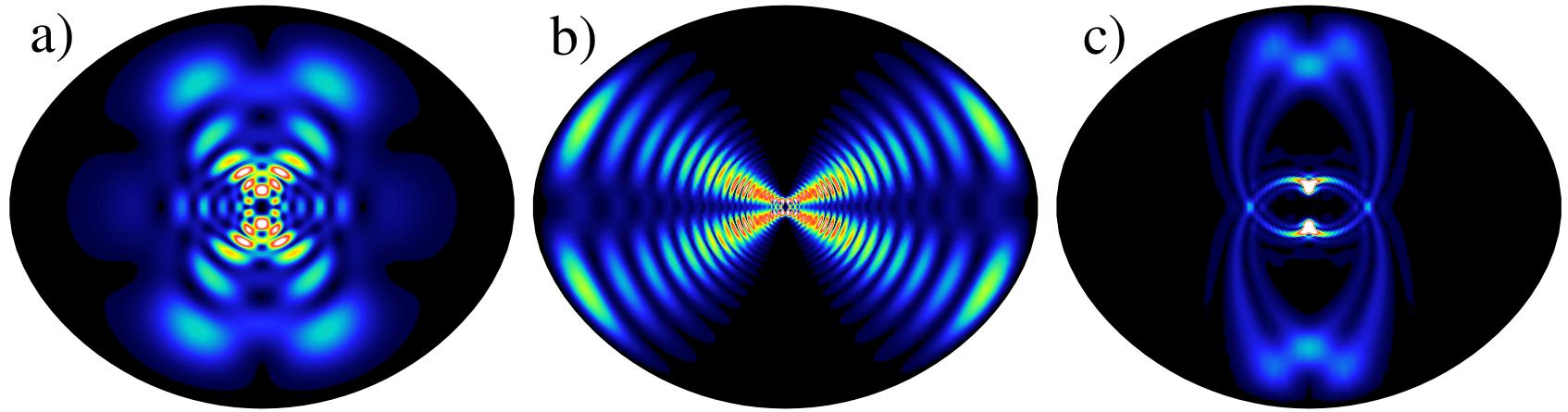}}
  \caption{Different classes of gravito-inertial modes. 
           Left to right: an unperturbed super-inertial mode, a sub-inertial
           mode and a rosette super-inertial mode.  The shown quantity is the
           mode kinetic energy. Figure taken from \citet{BLR13}.
\label{fig:geom_g}}
\end{figure}

Gravito-inertial modes, just like non-rotating gravity modes, are in the
low-frequency part of the oscillation spectra and propagate in stellar
radiative zones.  In early-type stars, they are likely to be excited in
$\gamma$Dor and SPB stars, in which they can provide insight on the conditions
near the stellar core.  Their restoring force is buoyancy, that combines with
the Coriolis acceleration. As the Coriolis acceleration scales with
$2\Omega/\omega$, its effect is more important for low-frequency modes in the
co-rotating frame.\\ Gravito-inertial modes are usually divided in three
geometries, depending on their frequencies and presented in
figure~\ref{fig:geom_g}.  We distinguish sub-inertial modes which propagate at
angular frequencies $\omega < 2\Omega$ and super-inertial modes at $\omega >
2\Omega$. 

The effect of rotation on sub-inertial modes is the appearance of a forbidden
region in which oscillations cannot propagate.  \citet{BLRR10} describes this
region in the case where the stratification dominates over rotation
($N\ll\Omega$) as a cone that extends at an angle
$\theta\sim\arccos(\omega/2\Omega)$ from the rotation axis.  This confirms the
equatorial confinement of the modes in rotating stars predicted by
\citet{matsuno66, LS90}.  The precise shape of the boundary of forbidden
regions for the propagation of sub-inertial modes has been refined to include
stratification \citep{DRV99} and differential rotation \citep{MBRB16}.  

Most super-inertial modes are marginally affected by rotation, as their
geometry stays close to their non-rotating equivalent, resembling simple
spherical harmonics with radial and horizontal scales of the same order of
magnitude at all rotation rates.

The most striking exception is rosette modes, that are super-inertial modes
whose geometry is significantly different from any spherical harmonic.  They
were described for the first time by \citet{ballot_etal12}.  These modes emerge
from a close degeneracy between modes whose frequencies would be near-identical
in the absence of rotation.  \citet{TS13} show that such degeneracies appear
for rotation rates as low as 10\% of the critical rotation rate and lead to
numerous families of rosette modes. This suggests that rosette modes are not a
rare occurrence in moderate rotators, even though their distribution in extreme
rotators is still unknown.  It appears that retrograde non-axisymmetric rosette
modes tend to form preferentially over their prograde counterparts
\citep{ST14}. It is also worth noting that because they originate from coupling
between different non-rotating modes, rosette modes cannot be found under the
traditional approximation.\\ These modes are spatially concentrated on very
thin layers around the stellar core and describe a rosette pattern aligned with
the rotation axis.  Ray tracing shows that their energy focuses on short closed
loops \citep{ballot_etal12, prat_etal16}, and that the period and orientation
of these closed loops can be related to the original family of modes
\citep[whose identification is otherwise lost,][]{Takata14}.  The surface
signature of rosette modes is still to be derived precisely and be identified
with observed modes, in which case they could provide useful constraints on a
star's deep layers.

\subsection{Other modes}
Finally, other classes of modes can exist in rapidly-rotating stars. 
Rossby modes are, among inertial oscillations, the ones that attracted most
interest from the astrophysical community.  Theoretical studies, notably by
\citet{PP78, saio82, saio18}, show they  are large-scale toroidal modes that
induce significant flows at the stellar surface that are less affected by
equatorial confinement at high rotation rates compared to gravito-inertial
modes.

Inertial modes can propagate in convective zones, restored by only Coriolis
forces. They propagate in the core of intermediate-mass stars in a frequency
domain that overlaps with gravito-inertial waves propagating elsewhere in the
star \citep{RGV01,BR13}.  The signature of such modes in the core of
$\gamma$Dor stars can be detected through their coupling with gravito-inertial
modes reaching the stellar surface \citep{ouazzani20, saio21}.  Similarly,
so-called overstable convective modes resonances have been suggested as an
excitation mechanism in rotating stars from $2\Msun$ upwards.  These modes are
core inertial modes, that couple with gravito-inertial modes in the envelope at
frequencies close to core rotation rate multiples \citep{osaki74, LS20,lee21,
lee22}.

Mixed modes also exist in rapidly-rotating intermediate-mass stars, as rapid
rotation extends the gravito-inertial domain to higher frequencies, making
coupling with pressure modes more likely \citep{osaki74}.  These modes are very
interesting as they probe both the deep layers of the star and carry a surface
signature, but their geometries and frequencies are affected in a complex way
by the coupling.  Such modes are detected in $\delta$Sct stars such as
Rasalhague \citep{monnier_etal10} or HD174966 (Garc{\'\i}a Hern{\'a}ndez,
private comm.), and become more likely with increasing age as the \BV frequency
in the star increases and allows for higher-frequency gravito-inertial modes
\citep{aerts_etal10}.  They have also been computed in two-dimensional models
by, e.g., \citet{mirouh17}.

\section{Regular patterns in oscillation spectra}
\label{sec:patterns}

Now that the mode geometries in rapidly-rotating stars have been introduced, I
will discuss the regular patterns these modes are expected to follow according
to theoretical calculations, and how the wealth of space-based measurements and
innovative pattern-recognition techniques have permitted their detection.

\subsection{Regular patterns in slow rotators}
Before delving into the frequency patterns of rapidly-rotating star spectra, it
is useful to remind the reader here of the regular patterns that can be found
in slowly-rotating stars with solar-like oscillations. I refer the reader to
the review of \citet{jackiewicz21} for more details.

Helioseismology has established that solar oscillation frequecies are
distributed on a regular comb.  Once modes are identified univocally using the
quantum numbers $(n, \ell, m)$, we find that high radial order modes present a
regular spacing in frequency $\nu$ for acoustic modes and in period $\Pi$ for
gravity modes. We define the p-mode large separation \citep{tass80} as
\begin{equation}
  \Delta\nu = \nu_{n+1, \ell} - \nu_{n, \ell} = \left( 2 \int^{R}_{0} \frac{{\rm d}r}{c_{\rm s}} \right)^{-1},
\end{equation}
where $c_{\rm s}$ is the speed of sound. The g-mode period spacing \citet{shib79} is defined as
\begin{equation}
  \Delta\Pi_\ell = \Pi_{n+1, \ell} - \Pi_{n, \ell} = \frac{2\pi^2}{\sqrt{\ell(\ell+1)}} \left(\int_{\rm g\ cavity} N \frac{{\rm d}r}{r} \right)^{-1}
\end{equation}
where $N$ is the \BV frequency that quantifies stratification.\\
In solar-like pulsators, in which mode excitation is due to convection and
stochastic, it is possible to define another seismic observable, $\nu_{\rm
max}$, the frequency at which the p-mode spectrum envelope reaches its maximum
amplitude. The readily-available quantities $\Delta \nu$ and $\nu_{\rm max}$
can then be linked to the stellar mass and radius via scaling laws
\citep{brown91, kallinger10, mosser2010}. Improving these relations has been a
significant part of the recent effort for those stars, to refine the derived
masses and radii, include new parameters or automate the process.

When the star rotates, the degeneracy in $m$ is lifted and non-radial modes
spawn multiplets as $m$ ranges from $-\ell$ to $\ell$.  The introduction of
rotation splittings for slow rotators, the frequencies of non-axisymmetric to
their axisymmetric counterpart through 
\begin{equation}
  \nu_{n,\ell,m} = \nu_{n,\ell} - \frac{m\Omega}{2\pi} \left(1-\mathcal{C}_{n.\ell}\right).
\end{equation}
$\mathcal{C}_{n.\ell}$ is the Ledoux constant, that tends asymptotically to
$1/(\ell(\ell+1))$ for gravito-inertial modes and to $0$ for pressure modes
\citep{ledoux51}.  More recently, rotational splittings have been shown to be
slightly different depending on the nature and propagation domain of the
underlying mode, thus allowing for the derivation of rotation profiles using
mixed modes
\citep[e.g.][]{beck2012,deheuvels2014,triana17,dimauro18,ahlborn20}.
Automating this analysis led to the development of inversion techniques, such
as RLS \citep[Regularized Least Squares,][]{jcd90} or SOLA \citep[Subtractive
Optimally Localised Averages,][]{sola}.  The ultimate aim is a characterisation
of the stellar rotation or structure throughout the star, as was done for the
Sun \citep{ked2011}.

As mentioned, the exploitation of regular spacings requires univocal mode
identification, which presents several challenges in rapidly-rotating
intermediate-mass pulsators.  First, because of Coriolis forces and centrifugal
distortion, frequencies are significantly shifted and both gravito-inertial and
pressure modes assume complex geometries. Because of the complexity of the
excitation mechanisms, there is no reliable way of predicting mode amplitudes
theoretically and the mode selection effects are not fully understood.

\subsection{Acoustic modes}
\begin{figure}
\centerline{ \includegraphics[width=0.7\textwidth]{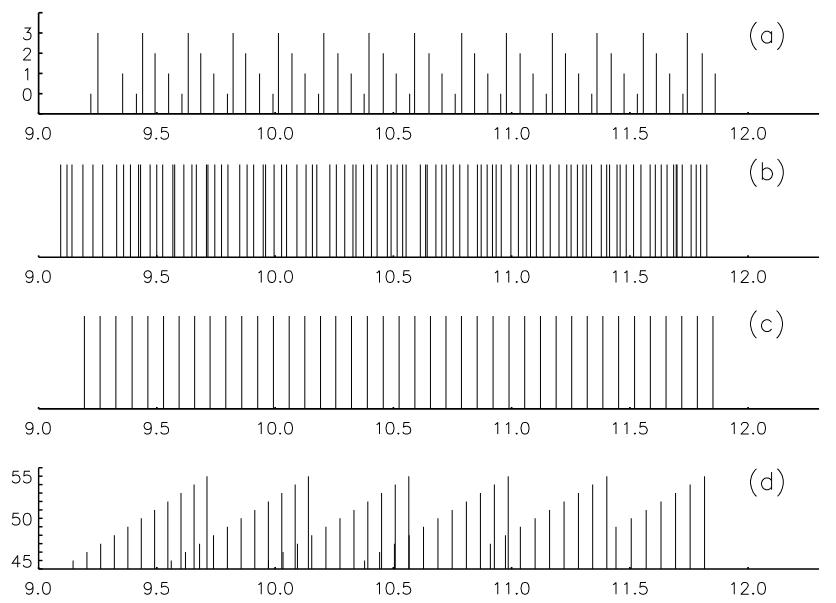}}
  \caption{Frequency sub-spectra for the four classes of axisymmetric acoustic
  modes.  (a) 2-period island modes, (b) chaotic modes, (c) 6-period island
  modes, (d) whispering gallery modes.  The horizontal axis is the mode
  frequency in units of $\omega_{\rm p}$ -- computed from the keplerian
  velocity computed using the polar radius --, and the vertical axis represents
  the second quantum number -- the rapid rotator equivalent to the degree
  $\ell$ -- when defined.  Figure taken from \citet{LG09}.
\label{fig:4classes}}
\end{figure}

The ray dynamics analysis of \citet{LG09} identified regular patterns for
several subclasses of acoustic modes.  There are four geometry classes for p
modes: 2- and 6-period island, whispering gallery, and chaotic modes. Modes in
each of these classes follow different patterns, as illustrated by the
schematic subspectra presented in figure~\ref{fig:4classes}. 

2-period island modes are the ones for which the most progress has been
achieved. They are expected to follow a linear distribution of the kind
\begin{equation}
  \label{eq:reg_2i}
  \nu_{\ntilde,\ltilde,m} = \ntilde \widetilde{\Delta\nu} + \ltilde \widetilde{\delta\nu} - m s_m(\Omega) + \tilde\alpha,
\end{equation}
where $\widetilde{\Delta\nu}$ is the island-mode large separation (and is expected
to be commensurate to half the non-rotating large separation $\Delta\nu$),
$s_m$ is the rotational splitting which depends on the rotation rate of the
propagation domain, and $\tilde\alpha$ is fixed so that the formula is exact at
a reference frequency.  This asymptotic relation has been studied extensively
both from ray theory \citep{LG08, LG09, pasek11, pasek12} and complete
calculations of the oscillations \citep{RLR08, mirouh_etal19, reese_etal21}.\\
Owing to the precision of asteroseismic measurements and the development of
dedicated techniques, such a regular frequency distribution has been detected
in observed spectra. \citet{AGH15} identified regular patterns in acoustic
spectra for a sample of eclipsing binaries and extracted a large separation
that they attributed to the likely visible 2-period island modes.  Using
eclipses and the binary orbits, they were able to link the large separation
with the mean stellar density through a scaling law.  Their scaling relations
were later confirmed from fully-2D models and oscillations by
\citet{mirouh_etal19}: in this work, the synthetic 2-period island modes were
sorted out by means of a convolutional neural network for a range of models at
various rotation rates and ages. It confirmed the scaling between mean density
and that this relation does not depend explicitly on the rotation rate.  This
scaling has also been recovered for low-mass $\delta$Sct stars using grids of
1D models with enhanced mixing by \citet{rodriguez20} and on the larger samples
of \citet{paparo16b, AGH17, bedding20}.  A further confirmation of the 2-period
island nature of the observed modes lies in the detection of half the large
separation in the pattern detection \citep{AGH09, ARB21}, which is an exclusive
feature of island modes \citep{evano_etal19}.  It is also worth noting that the
modes detected in $\delta$Sct stars are most often not in the asymptotic
regime, so that the measured large separations are about 15\% below their
asymptotic values \citep{AGH09, mirouh_etal19}.\\ Rotational splittings have
also been found in the 2-period island mode spectra of $\delta$Sct stars
\citep[e.g. ][]{paparo16b,ARB21}.  While rotational splittings in slow rotators
are symmetrical (that is, $\nu_{n,\ell,-m} - \nu_{n,\ell,0} = \nu_{n,\ell,0} -
\nu_{n,\ell,m}$), splittings in rapidly-rotating stars lose this property. They
also are often so large that asymmetric multiplets blend in the spectrum,
making them harder to identify and exploit. Modelling these splittings can rely
on perturbative treatments to the first-order \citep[e.g.][]{schou98,
deheuvels2014}, second-order {\citep[e.g.][]{saio81, suarez09} and third-order
\citep{SGD98, OG12} that model the asymmetry of the splittings.  A complete
calculation that applies to all rotation rates provides the generalised
rotational splittings, that is $\nu_{n,\ell,-m} - \nu_{n,\ell,m}$, through an
integral of the rotation rate over the modes' propagation domain and the
definition of rotational kernels \citep{RTMJSM09, reese_etal21}, thus paving
the way to inversion methods for rapid rotators.\\ Since the discovery of
island-mode regular patterns by \citet{AGH15}, several strategies have been
developed to extract large separations from observed spectra. Investigating the
distribution of frequency differences, computing Fourier transforms of the
spectrum itself, spectrum autocorrelation functions have all proved efficient
to derive both the large separation and rotational splittings
\citep{mantegazza12,reese_etal17,ARB21}.  A brand new approach based on the
entropy content of observed oscillation spectra offers another promising way of
extracting regular separations (Su\'arez, this volume).\\ To complement the
regular patterns discovered for 2-period island modes and their link with the
stellar density, \citet{SBF18, SBF20} have investigated the envelope of the
oscillation spectra of $\delta$Sct stars in search of another asteroseismic
indicator. They found that the peak frequency of this asymmetric envelope,
$\nu_{\rm max}$, can be related to the effective temperature and surface
gravity of the star. Assessing these two quantities for stars featuring gravity
darkening can yield an estimate of the stellar inclination, and in turn a
model-dependent estimate of the stellar mass and radii in a way that is
reminiscent of the approach used for solar-like stars.\\
Recently, \citet{salmon22} has developed a forward-modelling framework for
$\beta$ Cep stars, that relies on low-order pressure modes along with gravity
and mixed modes, to infer fundamental parameters. Their hare-and-hounds
exercise, which relies on 1D Geneva models \citep{eggenberger08}, yields
accurate parameters starting from only $\sim 5$ identified oscillation
frequencies.  Such a framework could serve as an example for the exploitation
of island modes computed from 2D models, using the associated theoretical tools
discussed above (see also section~\ref{sec:rasalhague}).

\begin{figure}
\centerline{ \includegraphics[width=0.7\textwidth]{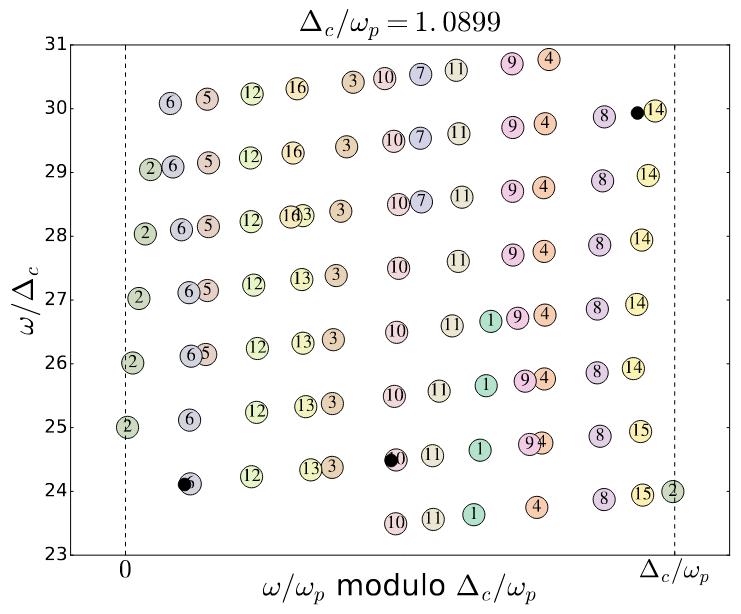}}
  \caption{\'Echelle diagram showing all chaotic modes in a polytrope in the
  high-frequency range -- the frequency unit $\omega_{\rm p}$ is computed
  from the keplerian velocity computed using the polar radius --. Series of
  similar modes are labelled 1 to 16, while black dots are modes that do not
  belong to any series.
  Figure taken from \citet{evano_etal19}.
\label{fig:echelle_chaot}}
\end{figure}

Frequency subspectra of chaotic modes are dense, and were originally thought
not to have structure \citep{LG09}. \citet{evano_etal19} computed the
autocorrelation of these subspectra and found evidence of a regular spacing.
They show that while chaotic modes do not tend to an exact asymptotic
distribution of oscillation frequencies, they exhibit a pseudo-large separation
$\Delta_{\rm c}\nu$ for all their models at all rotation rates.  They relate
$\Delta_{\rm c}\nu$ with the average acoustic time over the meridional plane
$T_0$ through $\Delta_{\rm c}\nu\sim2\pi/T_0$.  Chaotic modes thus form series
of modes that present a very similar geometry.  Figure~\ref{fig:echelle_chaot}
presents an {\'e}chelle diagram -- that is a frequency spectrum folded every
$\Delta_{\rm c}\nu$ to bring out patterns -- for chaotic modes in a polytrope.
Well-defined ridges follow the series, but some of them are interrupted at high
frequencies, and are replaced in the spectrum by another or by a series of
high-frequency island modes.  The pseudo-large separation of chaotic modes is
very close to that of island modes (within 1\%).  However, the detected
patterns are not expected to come from this spacing, as the half-large
separation detection is not expected from chaotic modes.  

Island modes of period 6 were identified as one of the four classes of acoustic
modes emerging from ray theory by \citet{LG09}. They propose a simple
asymptotic relation for these modes:
\begin{equation}
  \label{eq:reg_6i}
  \nu_{n', \ell', m} = n'\Delta'\nu + m s'_m + \alpha'
\end{equation}
where the various terms match are similar to those of equation (\ref{eq:reg_2i}).
The details of this asymptotic relation, and the link between the separation
$\Delta'_\nu$ and stellar structure quantities are still to be investigated.

Finally, whispering gallery modes also seem to follow regular patterns.
However, while it is straightforward to define suitable quantum numbers from
the energy distribution of whispering gallery modes, \citet{LG09} did not find
a satisfactory fit of their frequencies to a linear law in the form of
equations~(\ref{eq:reg_2i}) or (\ref{eq:reg_6i}).  More studies are required to
understand the nature of the patterns these modes follow.\\

\subsection{Gravito-inertial modes}
\label{sec:gip}
In the absence of rotation, the spectrum of high-order gravity modes is
understood as a set of modes regularly spaced in period, which can be
represented by a flat line in a $P-\Delta P$ diagram.  The value of the period
separation provides an estimate of the stellar age, as it decreases as a peak
in the \BV builds up at the core-envelope interface.  The deviation from this
flat distribution, in the form of with regular dips, is an indicator of a
chemical gradient at the core-envelope interface whose depth can be related
with mixing efficiency \citep{miglio08}. The pattern described by these dips
informs us about the location of the chemical gradient.

When the star rotates, the general trend in the $P-\Delta P$ plane is not
necessarily flat but varies monotonically \citep{bouabid2013, vanreeth2015}.
Using the traditional approximation, \citet{ouazz2017} thus defines the slope
of this trend as a new observable that is directly linked to the rotation rate
near the stellar core modulated by the azimuthal order $m$. These ridges were
observed in many $\gamma$Dor \citep[e.g.][]{bedding15, GL19, li2020b, li2020a,
garcia22} and SPB stars \citep[e.g]{papics2017, pedersen21,sz21,sz22}, with a
variety of behaviors related to mixing and rotation.  

Forward modelling has driven a lot of the progress on g-mode asteroseismology
of slowly to moderately rotating stars in the recent years. Through the
definition of rigorous frameworks \citep{aerts18, johnston19}, state-of-the-art
1D models and modes computed assuming the traditional approximation, partial
mode indentification has been obtained for numerous individual stars
\citep[e.g.][]{zhang18,szewczuk18}. This process yields not only stellar
fundamental parameters such as mass and age, but also constraints on
core-interface mixing \citep[e.g.]{wu20,pedersen21} or binary interactions
\citep{sekaran21, guo22}.\\
Recent works relying on the traditional approximation also provide
model-independent approaches. For instance, \citet{vanreeth16, li2020a,
takata20b, takata20} use the rotation frequency and the buoyancy radius as free
input parameters.  \citet{christophe18} suggests a clever rescaling of the
period spectra to extract signatures of differential rotation and buoyancy
glitches from the gravito-inertial mode spectrum. \\ \citet{christophe18} also
emphasizes the well-known limits of the traditional approximation rotation:
attempts at extending it to include radially differential rotation
\citep{mathis09} or centrifugal distortion \citep{dhouib21} may provide extra
insights on the physics of g-mode pulsators.  However, these approaches require
a prescription for a differential rotation profile (which is prone to a lot of
degeneracy) and are computationally expensive.

While the traditional approximation has led to a significant breakthrough in
the analysis of g-mode pulsators, permitting ensemble seismology and forward
seismic modelling for this class of stars, performing the full calculation of
gravito-inertial modes is the logical next step to include differential
rotation and enable the search for rosette modes in observations. 

\subsection{Rossby modes}
Rossby modes have long been theorized, and were first detected in rapid
rotators by \citet{vanreeth16}.  They have since been detected in a variety of
g-mode pulsators \citep[both $\gamma$Dor and SPB stars, e.g.][]{saio18, li19,
takata20}.  Their ubiquitous nature suggests an easy excitation mechanism, and
stellar activity or tidal forcing have been proposed.  Rossby modes present
spacings that appear in the $P - \Delta P$ plane, like gravito-inertial modes.
They present small period spacings, that increase rapidly with the period
\citep[leading to a characteristic upward slope, ][]{vanreeth16}.
\citet{saio18} suggests that the ``hump-and-spike'' stars discovered by
\citet{balona17} are spotted stars in which the ``hump'' is the Rossby-mode
spectrum and the ``spike'' a harmonic of the stellar rotation rate.  The
calculations underlying this description are made within the traditional
approximation framework, and have no equivalent based on complete
two-dimensional calculations.

\section{Line-profile variations}
\label{sec:lpv}

Modes that propagate in the outer layers of the star induce a small distortion
of the stellar surface along with temperature variations.  In rotating stars,
these oscillatory motions leave a Doppler signature in the spectral line
profile that can be tracked in time through a series of spectroscopic
observations.  Such line-profile variations (LPVs) are not simple to analyse as
they depend on the geometry of each oscillation mode present at the surface,
but their analysis can offer a partial mode identification. Once a specific
line profile is selected or an averaged one is computed, a spectral analysis of
its time variations yields oscillation frequencies and azimuthal orders.  

A lot of work has been done on the theoretical description of LPVs induced by
both pressure modes and gravito-inertial modes.  For instance, the modelling
efforts for p-mode LPVs \citet{balona87,cugier93,zima06} permitted the
interpretation of spectroscopic measurements in $\delta$ Sct and $\beta$ Cep
stars \citep[e.g.][]{aerts92,zima06b}. Theoretical developments based on the
perturbative approach \citep[e.g.][]{shutt21} or the
traditional approximation of rotation \citep[such as][]{LS90,townsend97}
yielded similar results to interpret $\gamma$ Dor pulsators.\\
Some of these methods have been implemented in widespread software packages.
For instance, the BRUCE and KYLIE packages \citet{townsend97} rely on the
traditional approximation for the calculation of theoretical LPVs of
gravito-inertial modes in moderate rotators \citep[as demonstrated
in][]{bowman22}.  Another widespread implementation is that of the Fourier
Parameter Fitting method (FPF) of \citet{zima06}. The FAMIAS code
\citep{famias} does not rely on the traditional approximation but on a
first-order perturbative calculation, which enables the analysis of LPVs due to
both gravito-inertial and pressure modes, but restricts it to slow rotators.

LPVs have also been detected in several rapidly-rotating stars
\citep[e.g.][]{TS98, balona01, Poretti09} and require special care. While the
theoretical developments of \citet{LS90, townsend97} are used for
gravito-inertial modes in rapid rotators, pressure-mode LPVs have benefitted
from two-dimensional codes, starting with \citet{clement94}.  The work of
\citet{reese_etal17b} is the most up-to-date such effort: it relies on
non-adiabatic modes computed with the high-resolution oscillation code TOP
\citep{reese_etal09,reese_etal21} paired with the self-consistent,
two-dimensional ESTER models \citep{REL13, RELP2016}.\\
\citet{reese_etal17b} shows that p-mode LPV signatures are maximal in the wings
of the line profile, contrary to what is expected in slow rotators
\citep{LS90}.  This can be explained by the complex geometries of modes in
rapidly-rotating stars, making LPVs equally more complex to decipher. The
centrifugal flattening and gravity darkening make the amplitude of the
signatures dependent on the stellar inclination, and different modes affect
different spectral lines in various ways, depending on the range of temperature
at the stellar surface \citep[see, e.g.][for the rapid rotator
Vega]{takeda2008}.\\
Figure~\ref{fig:lpv} presents the result of such calculations for a 2-period
island mode $9\Msun$ ESTER model at 50\% of its breakup velocity, This
calculation relies on a few simplifying assumptions such as Gaussian
equilibrium line profiles and rather crude limb- and gravity-darkening
prescriptions.  These assumptions will have to be relaxed to properly interpret
time series of high-resolution spectra, and will surely be the topic of
upcoming work. 

\begin{figure}
\centerline{ \includegraphics[width=0.7\textwidth]{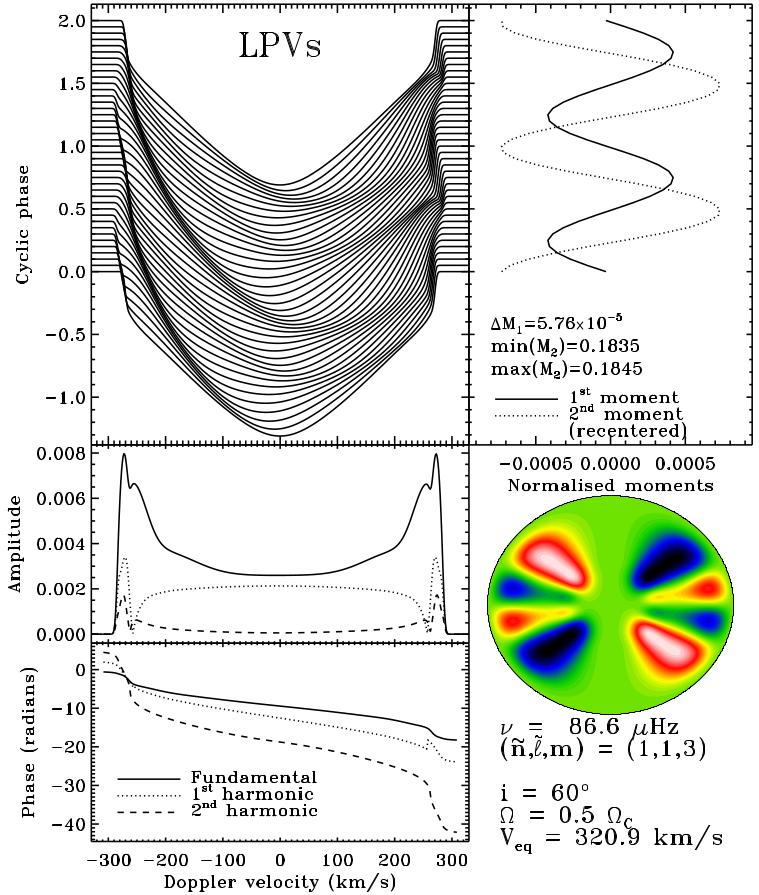}}
  \caption{Line-profile variations in a $9\Msun$ ESTER model at 50\% of its
  breakup velocity.  Left, top to bottom: stacked line profiles, amplitudes and
  phases of the first three harmonics.  Right, top to bottom: first and second
  moments of the LPVs, meridional cut of the oscillation mode, and model
  properties.  Figure taken from \citet{reese_etal17}.
\label{fig:lpv}}
\end{figure}

Eventually, adapting widespread automated procedures such as the FPF method
\citep{zima06} to two-dimensional models will enable a simpler detection of
identified modes that will serve as anchors for both pattern recognition and
forward modelling.

\section{Forward seismic modelling of fast rotators: the example of Rasalhague}
\label{sec:rasalhague}

Forward - or direct - modelling is a method of obtaining accurate properties of
stars by directly modifying its input parameters and computing associated
observables until a satisfactory match with measurements is found.  This
strategy is adapted for rapidly-rotating stars, especially when the number of
oscillation modes detected in them is relatively small and the regular patterns
in the spectra are difficult to disentangle. 

The necessary steps for forward asteroseismic modelling are :
\begin{enumerate}
  \item compute an appropriate model for a reasonable set of input parameters,
    for rotating stars this shall include centrifugal flattening and gravity
    darkening,
  \item establish criteria -- e.g. on mode visibility or excitation -- to
    select the most observable oscillation modes from the synthetic spectrum,
  \item identify the selected modes thus found with the observed ones,
  \item repeat the process by varying the initial model parameters until a good
    match is obtained.
\end{enumerate}
This strategy was successfully applied to slowly-rotating classical pulsators,
such as HD 129929 \citep{aerts_etal2003} for which 6 gravity-mode frequencies
were detected.  By means of non-adiabatic oscillations, \citet{aerts_etal2004,
dupret_etal04} not only obtained a stellar mass and age with uncertainties of a
few percent, but also constrained envelope convective penetration and
differential rotation.  Some other successes of the forward-modelling approach
for g-mode pulsators are given in section~\ref{sec:gip}.

I present here an application of the forward modelling approach to the
rapidly-rotating $\delta$ Scuti star Rasalhague ($\alpha$ Ophiuchi A,
HD159561). It is a rapidly-rotating pressure-mode pulsator, in which the
centrifugal distortion is expected to significantly impact oscillation
frequencies and geometries.

CHARA interferometry reveals that this A5 star is seen equator-on \citep[$v\sin
i = 239\pm12$\,\kms and $i = 87.5\pm 0.6^\circ$,][]{zhao_etal09}.  It also has
a known K6 companion \citep{cowley69, hinkley_etal11}, spectroscopic
measurements that provide chemical information \citep{EN03} and 57 oscillations
frequencies measured with MOST \citep[][including mostly pressure modes and a
few gravito-inertial or mixed modes]{monnier_etal10}. These numerous
observations make Rasalhague an optimal example for forward modelling.\\
Until recently, the masses derived from fitting the binary orbit of this system
($2.40^{+0.23}_{-0.37}\,\Msun$, \citealt{hinkley_etal11}) and fitting the
stellar flux with Yonsei-Yale models ($2.18\pm0.02\,\Msun$,
\citealt{monnier_etal10}) covered a wide range. 
The precision of the binary orbit fit was significantly improved by observations near periastron
passage, yielding a mass of $2.20\pm0.06\Msun$, matching estimates from interferometry and models closely \citep{gardner21}.
The oscillation spectrum
contains 57 frequencies ranging from $1.7$ to $48.5$ cycles per day -- $20$ to
$560\,\mu$Hz. It contains both gravity and pressure modes over the same
frequency range, making it a $\delta$ Scuti/$\gamma$ Doradus hybrid pulsator.
The overlap is undeniably due to the star's rapid rotation and its impact on
mode frequencies, as slowly-rotating hybrid pulsators usually present a clearer
frequency separation between the p- and g-mode subspectra. 

\begin{figure}
\centerline{ \includegraphics[width=0.7\textwidth]{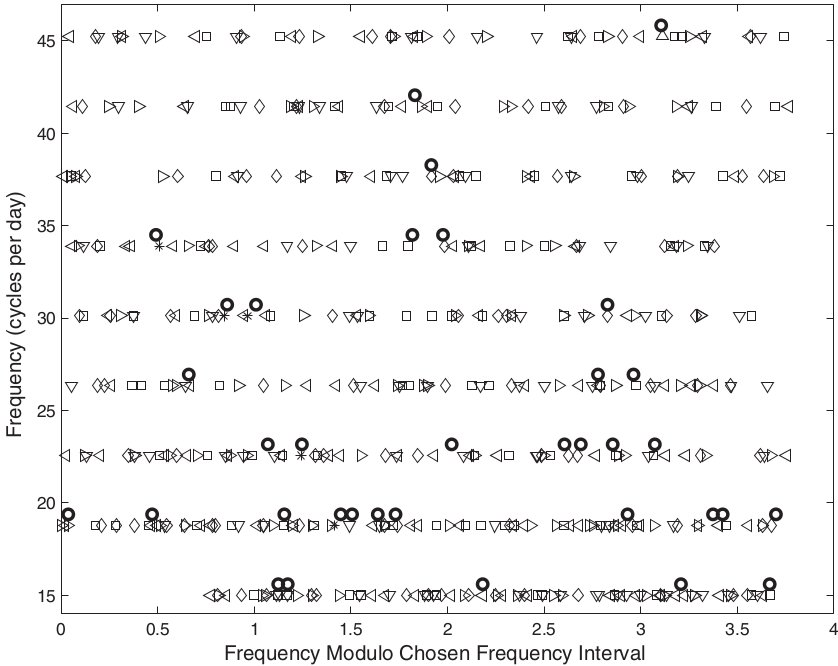}}
  \caption{Echelle diagram comparing synthetic modes at different azimuthal
  orders $m$ (squares: $|m|=0$, diamonds $|m|=1$, triangles $|m|=2$ and
  right-facing triangles $|m|=3$) with observed frequencies (thick circles
  artifically offset). Figure taken from \citet{deupree11b}.
\label{fig:ras_deupree}}
\end{figure}

The first attempt at modelling Rasalhague was by \citet{deupree11b}, using the
2D ROTORC code \citep{Deupree90, Deupree95}. He used nine solid-body rotating
and one slightly differentially-rotating models, and computes adiabatic
oscillations with $-4 \leq m \leq 4$, using six or eight spherical harmonics
for each mode.  He then identifies oscillation modes by matching MOST
frequencies to their closest synthetic counterpart within one stardard
deviation. Figure~\ref{fig:ras_deupree} presents the match between synthetic
frequencies for different azimuthal orders and the observed ones. The \'echelle
diagram is folded with a $\Delta\nu = 47.6\mu$Hz that emerges from the models,
but does not match the estimate of \citet{AGH15}.\\ While an interesting first
attempt, this study has clear limitations already mentioned in the original
paper. First, the synthetic spectrum computed by \citet{deupree11b} is so dense
that several synthetic frequencies lie within the observational uncertainties
for numerous modes, which makes the differentiation between models difficult.
More importantly, increasing the number of spherical harmonics to describe each
mode from six to eight leads to a drastic change in the mode geometry. This
shows that 6- and 8-harmonic calculations are underresolved, which undermines
the whole analysis.\\
The MOST dataset has been studied by \citet{AGH15, AGH17} who found regular
patterns in the oscillation spectrum. For frequencies above $116\,\mu$Hz they
found a large separation of $38\pm1\,\mu$Hz that corresponds to a mean density
of $0.123\pm0.021 \rho_\odot$. The $116\,\mu$Hz frequency is that of the
lowest-order radial pressure mode, while the $38\pm1\,\mu$Hz pattern is due to
2-period island modes. These seismic parameters and partial mode identification
has guided the new modelling effort I describe here.

This new calculation relies on the ESTER and TOP codes and is presented in
\citet{mirouh_etal14a, mirouh17}.  The ESTER code computes the structure of a
rotating star in two dimensions, including centrifugal distortion, gravity
darkening and meridional circulation, but treats mixing in a very simplified
manner that does not track core recession and cannot include a surface
convection zone. It is paired with the TOP code, a high-resolution spectral
code to compute the adiabatic and non-adiabatic oscillations of rotating
models.  For this work, we compute a $2.22\Msun$ model whose surface
temperature, radii and rotation match the interferometric measurements.  We
compute adiabatic modes with  $-4 \leq m \leq 4$, this time using twenty
spherical harmonics for each mode. To select the best candidates for
identification, we compute mode visibilities \citep[that is, the induced
surface signature][]{reese_etal13, reese_etal17} and the growth or damping rate
(that is, the rate at which the mode is excited or damped in the star).  Growth
and damping rates are directly obtained through solving the non-adiabatic
oscillation equations or from work integrals computed in the adiabatic
approximation, that is, assuming that non-adiabatic effects do not impact the
mode geometry \citep{unno_etal89, mirouh_etal14a}.

\begin{figure}
\centerline{ \includegraphics[width=0.7\textwidth]{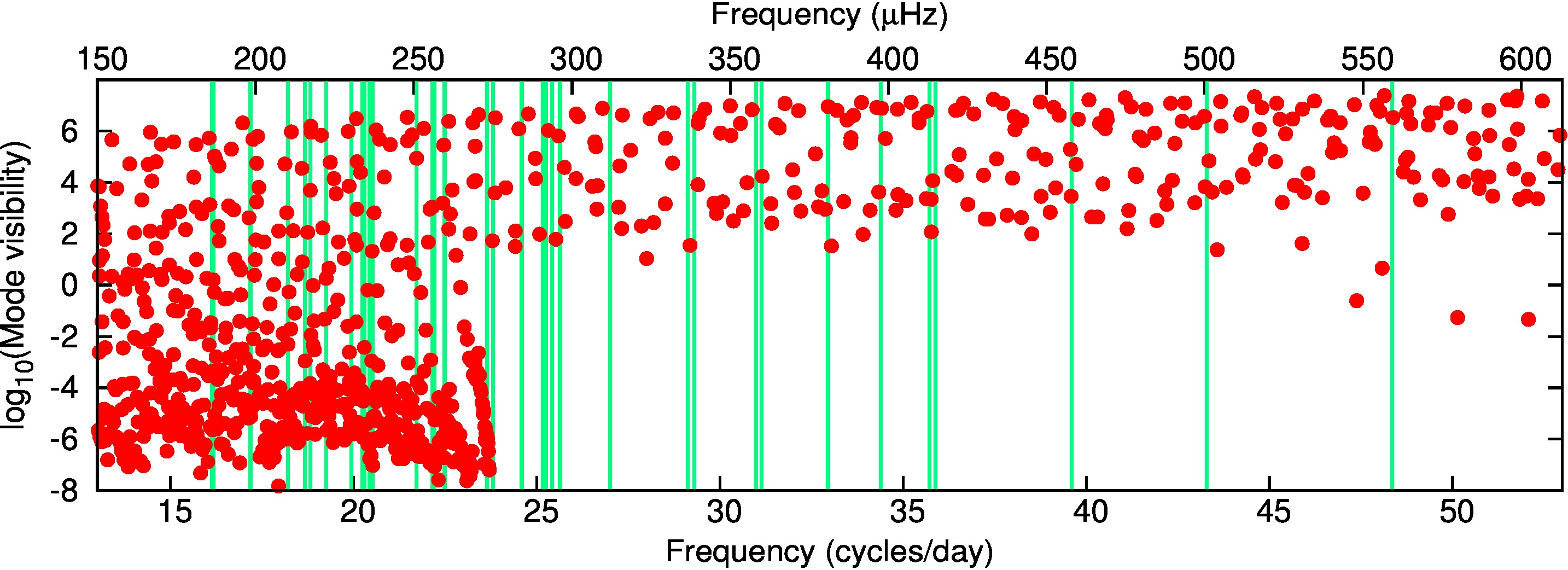}}
\centerline{ \includegraphics[width=0.7\textwidth]{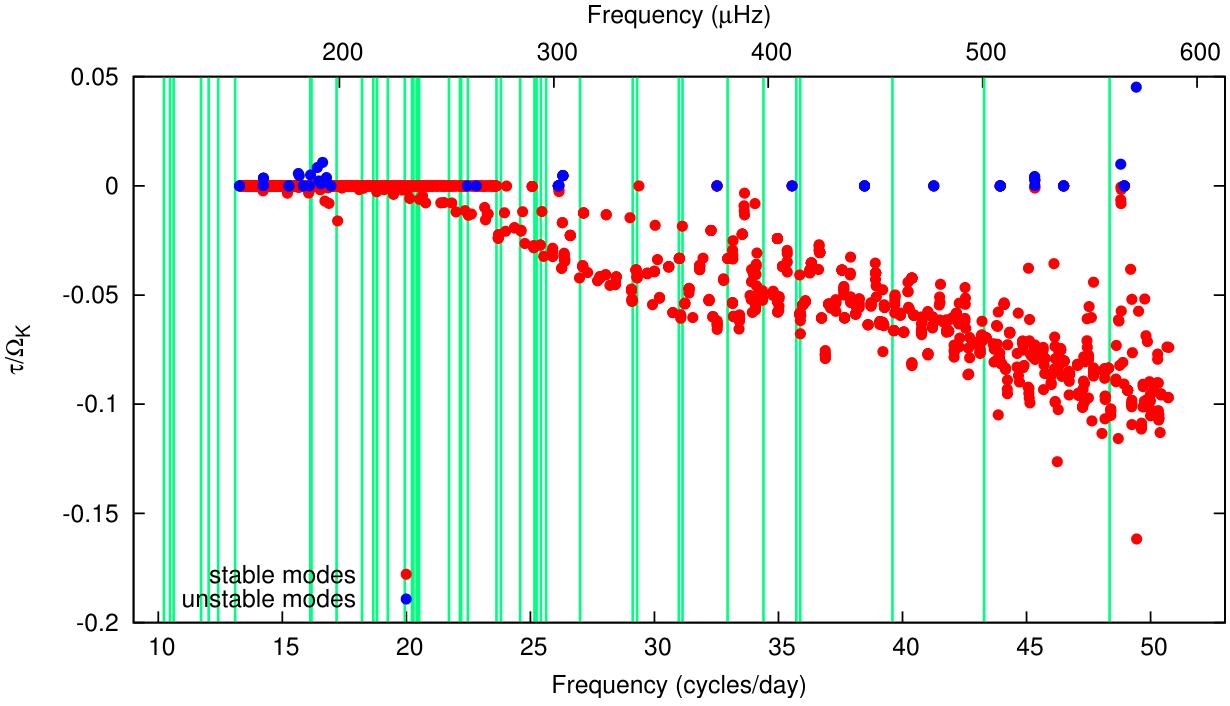}}
  \caption{Properties of axisymmetric ($m = 0$) modes.  Top: mode visibilities.
  Bottom: mode damping (red) or growth (blue) rates obtained from non-adiabatic
  calculations.  Green lines denote frequencies from \citet{monnier_etal10}.
  Figure taken from \citet{mirouh17}.
\label{fig:ras_mirouh}}
\end{figure}

The aim is to select the most visible excited modes that match the measured
frequencies. The top panel of figure~\ref{fig:ras_mirouh} shows the visibility
of each mode, and visual inspection brings out two overlapping populations.
Pressure modes are the most visible and cover the whole frequency range, while
gravito-inertial modes exist at low frequency ($\omega <\sqrt{N^2 +
4\Omega^2}$) and have lower visibilities.  This difference comes from the
energy distribution in the star, as gravito-inertial modes are focused around
the stellar core and seldom reach the stellar surface and acoustic modes
propagate up to the surface.

Computing growth and damping rates using work integrals in the quasi-adiabatic
approximation yields only linearly-stable modes, with acoustic modes being more
damped than gravito-inertial modes. This can be attributed to a poor
description of the modes at the stellar surface and prevents any identification
of the observed modes \citep{mirouh_etal14a}.  \\ Non-adiabatic calculations
reveal the presence of linearly-unstable modes \citep{mirouh17}, as shown in
the bottom plot of figure~\ref{fig:ras_mirouh}.  However, those modes do not
seem to match the expected distribution: most unstable modes are gravity modes
that describe a somewhat regular spacing which does not match the large
separation of \citet{AGH15,AGH17}, while the most visible modes are damped.
The origin of unstable (super-inertial) gravito-inertial modes in ESTER models
of Rasalhague comes as a surprise: as the models do not feature a convective
envelope, they cannot include the convective blocking excitation mechanism.  If
the computed modes are to explain the gravity modes observed by
\citet{monnier_etal10}, their excitation must be provided by another mechanism.
Candidates include the $\kappa$ mechanism for gravity modes, as proposed by
\citet{xiong2016}, the overstable convective modes suggested notably in
\citet{lee21}, or the elusive differential rotation mechanism suggested by
\citet{MBRB16}.

At this point it is thus impossible to identify modes. This may be due to the
non-adiabatic calculations whose stability must be ensured, and ESTER models'
inclusion of surface effects. A better understanding of the intertwined regular
patterns in the spectra (described in section~\ref{sec:patterns}) can also be
obtained by the systematic application of classification algorithms such as
that of \citet{mirouh_etal19}. This algorithm is a supervised machine learning
approach, that relies on a bank of well-identified pressure perturbation
profiles to separate rapidly a set of computed modes into the various
geometries presented in section~\ref{sec:geometries}.  \citet{mirouh_etal19}
focused on extracting patterns for island modes, and can be extended to study
other pressure and gravito-inertial modes.  Other machine learning approaches,
both supervised and unsupervised, applied to individual modes or entire
observed spectra, will undoubtedly be the next step towards mode identification
in rapid rotators such as Rasalhague. Rasalhague also seems to be a good
candidate for line-profile variation detections (described in
section~\ref{sec:lpv}), as it is seen equator-on which would maximize the
signature of island modes.  

These improvements will be necessary to obtain a satisfactory two-dimensional
seismic model of Rasalhague and open the way to the study of other
rapidly-rotating pulsators observed through interferometry, such as
Alderamin, Regulus or Achernar. Another very interesting star is the rapid rotator
Altair whose realistic modelling, albeit not seismic, was achieved by 
\citet{bouchaud20} and which features 15 oscillation frequencies \citep{ledizes21}.

\section{Summary}
\label{sec:ccl}
Rapid rotation makes early-type main-sequence pulsators a much more complex
object of study compared to slowly-rotating late-type counterparts.  While our
understanding of these stars has been lagging behind that of solar-like
pulsators for years, the gap is closing at an increasing speed.  The
application of the traditional approximation for gravito-inertial modes in
moderate rotators, and the development of two-dimensional models and
oscillation codes {for both pressure and gravito-inertial modes at all rotation
rates}, have unveiled the complex geometries and patterns the modes assume.
Pattern analysis in high-quality oscillation spectra reveal an increasing
amount of information on rotating stars.  I presented the topology of both
gravito-inertial and acoustic modes under the combined effect of the Coriolis
force and centrifugal distortion.

Kepler and TESS revealed numerous regularities in the low-frequency spectra of
$\gamma$ Doradus and SPB stars, that were matched by gravito-inertial and
Rossby modes computed with the traditional approximation. This opened a window
on differential rotation and mixing inside those stars, and statistics are now
building up to reach a better understanding of the structure and evolution of
these stars.\\ Pressure-mode pulsators such as $\delta$ Scuti benefitted from
recent progress in two-dimensional models and oscillation codes. Theoretical
patterns are identified and linked to structure quantities, and an entire
toolkit is now available to compute these modes, their visibilities and damping
rates. I described in this article two developments that may be the final
pieces of the puzzle.  One of them is the development of classification
algorithms to automate mode identification and derive patterns for each
subclass of acoustic modes, and the other is line-profile variations adapted to
rapidly-rotating stars to provide solid anchor points to forward seismic
modelling approaches.

Just like Kepler and TESS have allowed unprecedented progress in the
deciphering of classical pulsator oscillation spectra, the future PLATO mission
holds the promise of yet another revolution. The sheer amount of stars for
which lightcurves will eventually be available will be a stepping stone towards
a full understanding of rotation, angular momentum mixing and their impact on
intermediate- and high-mass stellar evolution.

\bibliographystyle{frontiersinSCNS_ENG_HUMS}
\bibliography{bibnew2}

\section*{Author Contributions}
GMM has written this entire article and is accountable for its content.

\section*{Funding}
GMM ackowledges support by ``Contribution of the UGR to the PLATO2.0 space
mission. Phases C / D-1'', funded by MCNI/AEI/PID2019-107061GB-C64.

\section*{Acknowledgments}
The author thanks Daniel R. Reese, Antonio Garc{\'\i}a Hern{\'a}ndez and Juan
Carlos Su{\'a}rez for insightful discussions that helped define and improve
this review, along with both anonymous referees whose detailed comments led
to significant improvements to this article. Special acknowledgements also go
to the PIMMS workshop attendees for the interesting discussions that took place
there.

\section*{Data Availability Statement}
Information about the datasets mentioned in this review can be found in the relevant publications.

\end{document}